
\tolerance=1000
\def\vs{{\bf s}}
\def\ket#1{{\vert{#1}\rangle}}
\def\bra#1{{\langle{#1}\vert}}
\def\cs{{\cal S}}
\newcount\eqnumber
\eqnumber=0
\def\en{\global\advance\eqnumber by 1
          \eqno(1.\the\eqnumber)}
\hsize 6.25 truein
\vsize 9 truein
\pageno=1

\vglue 2 \baselineskip
\centerline{\bf A Microscopic Approach to the Dimerization in}
\centerline{\bf the Frustrated Spin-$1\over2$ Antiferromagnetic Chains}
\vskip 2 \baselineskip
\centerline{Y. Xian}
\vskip 2 \baselineskip
\centerline{\sl Department of Mathematics, UMIST}
\centerline{\sl (University of Manchester Institute of Science and
                Technology)}
\centerline{\sl P.O. Box 88, Manchester M60 1QD, UK}
\vskip 4 \baselineskip

\noindent{\bf Abstract}. The spontaneous dimerization of the frustrated
spin-$1\over2$ antiferromagnetic chains is studied by a microscopic
approach based on a proper set of composite operators (i.e.,
pseudo-spin operators). Two approximation schemes are developed.
Firstly, a spin-wave approximation is made by a Dyson-Mal\'eev-like
boson transformation. The ground-state properties and the triplet
excitation spectra are obtained as functions of the coupling
parameter. Secondly, based on the pseudo-spin operators, a
microscopic treatment is formulated within the framework of the
powerful, systematic coupled-cluster method (CCM). Comparison between
various approximations is made. The advantage of the CCM for the
purposes of systematic improvement is emphasized.

\vskip 4 \baselineskip

\vfill
\eject

\centerline{\bf I. Introduction}
\nobreak\vskip\baselineskip \nobreak

Spontaneous dimerization of theoretical spin-lattice models was
perhaps first discovered by Majumdar and Ghosh [1] in 1969. They found
that for the one-dimensional (1D) spin-$1\over2$ system with
nearest-neighbour and next-nearest-neighbour couplings, the perfect
dimer state, in which every two adjacent atoms form a spin singlet
valence-bond, is the exact ground state of the Hamiltonian at a
particular coupling. Obviously, the translational symmetry is
spontaneously broken in the dimerized system, and the corresponding
ground state of the 1D model is doubly degenerate.

A spin-1 chain or other higher-order spin chains can also exhibit
dimerized valence-bond structures in their ground states [2]. Very
recently, a series of 1D $SU(n)$ spin-$s$ ($n=2s+1$) antiferromagnetic
models have been solved by Bethe {\it ansatz} [3], and the
corresponding dimerization order parameters have been exactly
calculated by the author [4]. In addition to the dimerized spin
chains, it is possible that some 1D spin systems (with integer spin
quantum numbers) favour the trimer configuration which is produced by a
sequence of spin-singlet states formed from every three adjacent
spins. Furthermore, dimerization or trimerization may also occur in
two or higher-order spatial dimensions. Clearly, a dimerized or
trimerized spin system can be viewed as a kind of solid in which the
corresponding simple valence-bonds are localized and the translational
symmetry is broken. The perfect dimer or trimer state is not in
general the exact ground state of a given Hamiltonian, but for some
systems the ground state may still possess a nonzero solid-like dimer
or trimer long-range order and hence show the characteristics of a
quantum solid.  (It therefore seems more appropriate to call these
solid-like systems collectively as `valence-bond lattices' [5].) Due
to quantum correlations, one expects that the corresponding long-range
order of those quantum solids will be reduced or vanish at certain
coupling strengths. A good example is provided by the ground state of
the spin-1 $SU(3)$ antiferromagnetic chain [4], where the dimerization
order parameter is reduced to 42\%.

One is quite familiar with phonons in ordinary atomic solids and
magnons in ferromagnets or antiferromagnets. In particular, the
spin-wave theory of Anderson [6] provides a simple and excellent
description of the spin correlations in the ground and low-lying
excited states for a number of antiferromagnets. Similar
approximations have also been developed for some dimerized systems.
Parkinson [7] formulated a spin-wave approximation based on a
spin-$1\over2$ dimer state. He focused on the spin-$1\over2$
Heisenberg model and employed the method of equation-of-motions. A
triplet excitation spectrum of $\sqrt2\sin k$ was obtained. This compares
well with the exact result of triplet spectrum of ${\pi\over2}\sin
k$. Chubukov [8] later provided a similar spin-wave theory by using a
transformation similar to that of Holstein-Primakoff to study
specifically the dimerization of the spin-1 chains with Heisenberg and
biquadratic exchanges. Read and Sachdev [9] investigated the
dimerization problem within the framework of Schwinger boson field
theory. Very recently, the author [10] has extended Parkinson's
theory to discuss possible trimerization of a 1D isotropic spin-1
system. The concept of dimerization has also been extended to the 2D
spin systems, e.g., the $J_1$-$J_2$ Heisenberg model on the square
lattice [11], etc.

In this article, I intend to investigate the dimerization of spin
systems by a systematic, microscopic approach.  Because of its
simplicity, I focus on the 1D spin-$1\over2$ Heisenberg chains with
nearest-neighbour and next-nearest-neighbour couplings, for which the
dimer state is the ground state at a particular coupling [1].
Following Parkinson [7], I study the dimerization in terms of a proper
set of composite operators (pseudo-spin operators). I extend and
reformulate Parkinson's theory so that the ground-state properties as
well as the excitations can be investigated. Firstly, by employing a
Dyson-Mal\'eev-like boson transformation [12] for the pseudo-spin
operators, a spin-wave theory is developed. The advantage of using
Dyson-Mal\'eev transformation lies in the fact that Hamiltonian can be
expressed in a compact form as a finite order polynomial of boson
operators, rather than as an infinite series which is the case when
the Holstein-Primakoff-like transformation is used (e.g. Ref.~[6]).
Secondly, I apply the powerful, systematic, microscopic many-body
theory of the coupled-cluster method (CCM) [13] based on the
pseudo-spin operators of the frustrated spin-$1\over2$ chain. A
systematic approximation scheme within the CCM approach is developed
for the ground state. The CCM has recently been successfully
applied to the various spin systems with Ising-like long-range
order [14] or with planar long-range order [15]. The excellent
results produced by the CCM approximations, particularly for the spin
systems with Ising-like long-range order, is the main motivation for
the current CCM approach to the dimerization problems.

It should be emphasized that the dimerized or trimerized states are not
merely mathematical artifacts. In fact, the 1D frustrated spin-$1\over2$
system has recently been shown to be relevant to the reconstruction of
fcc metal surfaces at finite temperature, with the spin dimerized
phase corresponding to the disordered flat surfaces [16].

The plan of this paper is as follows. In Sec.~II the composite
operators in a matrix representation are introduced and the
corresponding boson transformations are given. Sec.~III is devoted to
the spin-wave approximation for the 1D spin-$1\over2$ model. The
ground-state energy and excitation spectra are obtained as functions
of the coupling constants. In Sec.~IV I describe in detail the
microscopic CCM for the ground state of the dimerized spin system. The
results from the CCM approximation are compared with the spin-wave
theory. I conclude this article by a general discussion in Sec.~V.

\vskip \baselineskip

\newcount\eqnumber
\eqnumber=0
\def\en{\global\advance\eqnumber by 1
          \eqno(2.\the\eqnumber)}

\centerline{\bf II. Pseudo-Spin Operators and Their Bosonizations}
\nobreak\vskip\baselineskip\nobreak

I first consider a two-spin system, each with spin $1\over2$. For
completeness, some of the analysis by Parkinson [7] is repeated here.
Clearly, there are four states for such a two-spin system. If notations
$\ket\uparrow$ and $\ket\downarrow$ are used to represent spin up and down
states respectively, the singlet state can be written as
 $$
	\ket0={1\over\sqrt2}(\ket{\uparrow\downarrow}-
        \ket{\downarrow\uparrow}),	\en
 $$
and the triplet states with $s_{\rm total}^z\ (\equiv s^z_1+s^z_2)\ =
1,0,-1$ are given by respectively
 $$
	\ket1=\ket{\uparrow\uparrow},\hskip.3truein
	\ket2={1\over\sqrt2}(\ket{\uparrow\downarrow}
        +\ket{\downarrow\uparrow}),	\hskip.3truein
	\ket3=\ket{\downarrow\downarrow}. \en
 $$

Following Parkinson [7], I employ a matrix representation. Each of the
four states of Eqs.~(2.1) and (2.2) is then represented by a column
matrix with a single nonzero element. One can then introduce operators
$A_{ij}$ as having only a single non-zero element of a ($4\times4$)
matrix, namely $\bra{i'}A_{ij}\ket{j'}=\delta_{ii'}\delta_{jj'}$.
For example, $A_{10}$ ($A_{30}$) is an operator which increases
(decreases) $s^z_{\rm total}$ by one unit, while $A_{20}$ leaves it
unchanged. Their hermitian conjugates (i.e., transpose matrices) have
the opposite effects.  Together with other operators, these sixteen
operators form a complete set for the spin pair and any operator of
the pair can be written as a linear combination of these sixteen.

For a pair of spins, each of which has spin greater than one half,
similar operators can be defined. For example, there are nine states
for a two-spin system each with spin 1, and hence there are eighty-one
($9\times9$) $A_{ij}$ operators which form a complete set for a pair
of spin-1 atoms. For three-spin system each with spin 1, the
dimensionality of the matrix is twenty-seven [10].

It is worth pointing out that these $A_{ij}$ operators are {\it nonlinear}
in terms of the original single spin operators, for example,
 $$
  A_{00}={1\over4}-\vs_1\cdot\vs_2,\ \ \ \
  A_{01}=-{1\over\sqrt2}(s^+_1-s^+_2)(s^z_1+s^z_2),\en
 $$
etc. It is in this sense that I have referred to these $A_{ij}$ as {\it
composite operators} [10]. (Notice that $A_{00}$ is the usual spin-singlet
projection operator.)  Furthermore, it is easy to prove that they obey
the following pseudo-spin algebra,
$$ 	[A_{ij},\ A_{kl}]=A_{il}\delta_{jk} - A_{kj}\delta_{li}. \en $$
Therefore, I also refer to them as {\it pseudo-spin operators}. My
assumption in this paper is that it is more natural to study the
dimerization in terms of these composite operators rather than the
original single spin operators. The spin-wave theory in Sec.~III and
the CCM approximations in Sec.~IV are developed based on this
assumption.

In a straightforward manner, one can express the single spin operators
in terms of $A_{ij}$ operators [7]. They are given by [17]
 $$ \eqalignno{
	s^z_1&={1\over2}(A_{02}+A_{20}+A_{11}-A_{33}),\hskip.2truein
	 s^z_2={1\over2}(-A_{02}-A_{20}+A_{11}-A_{33}), &(2.5a)\cr
	s^+_1&={1\over\sqrt2}(A_{30}-A_{01}+A_{21}+A_{32}),\hskip.2truein
	 s^+_2={1\over\sqrt2}(A_{01}-A_{30}+A_{21}+A_{32}), &(2.5b)\cr
	s^-_1&={1\over\sqrt2}(A_{03}-A_{10}+A_{12}+A_{23}),\hskip.2truein
	 s^-_2={1\over\sqrt2}(A_{10}-A_{03}+A_{12}+A_{23}), &(2.5c)\cr }
 $$ \advance\eqnumber by 1
Recognizing that $A_{ij}$ obeys the pseudospin algebra of Eq.~(2.4), one can
make the following Dyson-Mal\'eev transformation [12],
 $$ \eqalign{
     A_{00}&=1-a_1^+a_1-a^+_2a_2-a^+_3a_3;\cr
     A_{p0}&=a^+_pA_{00}, \ \ \ \ \ \ A_{0p}=a_p;\cr
     A_{pp}&=a^+_pa_p, \ \ \ \ \ \ A_{pq}=a^+_pa_q,\cr} \en
 $$
where $p,q = 1,2,3$, and $a_p,a^+_p$ are three sets of boson
operators, obeying the usual boson commutation
 $$ [a_p,\ a_p^+] =1, \en $$
and with all other commutators yielding zero.

By definition the singlet state $\ket 0$ of Eq.~(2.1) is the vacuum
state of the bosons, namely,
 $$ a_p \ket 0 = 0, \ \ \ \ \ \ p=1,2,3. \en $$
The physical states correspond to the vacuum state $\ket 0$ and the
three states with only one boson excited. Furthermore, as the matrix
elements between physical and unphysical subspaces are equal to zero,
the transformation given by Eqs.~(2.6) is exact at zero temperature
just as in the case of the conventional spin-wave theory [6].

A general spin-$1\over2$ Hamiltonian can be expressed in terms of
$A_{ij}$ operators by Eqs.~(2.5) and then in terms of those three sets
of boson operators according to Eqs.~(2.6). In Sec.~III I consider a
spin-wave theory for the frustrated 1D model using these pseudo-spin
operators and their bosonizations. In Sec.~IV I develop a microscopic
formalism within the framework of the CCM also based on these
pseudo-spin operators.

\vskip \baselineskip

\newcount\eqnumber
\eqnumber=0
\def\en{\global\advance\eqnumber by 1
          \eqno(3.\the\eqnumber)}

\centerline{\bf III. Spin-Wave Theory}
\nobreak\vskip\baselineskip\nobreak

The 1D spin-$1\over2$ isotropic model with nearest-neighbour and
next-nearest-neighbour couplings is described by the Hamiltonian,
$$ H=\sum_{l=1}^N(\vs_l\cdot\vs_{l+1}+J \vs_l\cdot\vs_{l+2}), \en$$
where $J$ is the coupling constant and $N$ is the total number of
spins. I use the periodic boundary condition and choose even $N$ for
convenience. I have also taken the lattice spacing to be unity. At
$J={1\over2}$, the Hamiltonian of Eq.~(3.1) becomes the well-known
Majumdar-Ghosh model [1], which has two degenerate dimer
ground-states, with one of them given by
 $$ \ket D = \prod_{r=1}^{N/2}\ket 0_{2r-1,2r},
      \en$$
where the notation $\ket 0_{i,j}$ represents the singlet-paired state of
Eq.~(2.1). This dimer state $\ket D$ is shown graphically in Fig.~1.
After choosing the dimer indices as shown in Fig.~1, the Hamiltonian can
be written as
 $$ H = \sum_{r=1}^{N/2} \bigl[\vs_1(r)\cdot\vs_2(r) +
        \vs_2(r)\cdot\vs_1(r+1) + J\vs_1(r)\cdot\vs_1(r+1) +
        J\vs_2(r)\cdot\vs_2(r+1)\bigr]. \en$$
One can then express $H$ in terms of the composite operators $A_{ij}$ by
Eqs.~(2.5) as
$$ H=\sum_{r=1}^{N/2}(H_r+{1\over4}H_{r,r+1}+{1\over2}JH'_{r,r+1}), \en$$
with
$$ \eqalignno{
   H_r &\equiv {1\over4}-A^r_{00}, &(3.5a)\cr
   H_{r,r+1}&\equiv
    (-A^r_{02}+A^r_{11}-A^r_{20}-A^r_{33})
             ( A^{r+1}_{02}+A^{r+1}_{11}+A^{r+1}_{20}-A^{r+1}_{33})\cr
  &+( A^r_{01}+A^r_{21}-A^r_{30}+A^r_{32})
             ( A^{r+1}_{03}-A^{r+1}_{10}+A^{r+1}_{12}+A^{r+1}_{23})\cr
  &+(-A^r_{03}+A^r_{10}+A^r_{12}+A^r_{23})
             (-A^{r+1}_{01}+A^{r+1}_{21}+A^{r+1}_{30}+A^{r+1}_{32}),
    &(3.5b)\cr
  H'_{r,r+1}&\equiv
    (A^r_{02}+A^r_{20})(A^{r+1}_{02}+A^{r+1}_{20})
   +(A^r_{11}-A^r_{33})(A^{r+1}_{11}-A^{r+1}_{33})\cr
  &+(A^r_{01}-A^r_{30})(A^{r+1}_{10}-A^{r+1}_{03})
   +(A^r_{21}+A^r_{32})(A^{r+1}_{12}+A^{r+1}_{23})\cr
  &+(A^r_{10}-A^r_{03})(A^{r+1}_{01}-A^{r+1}_{30})
   +(A^r_{12}+A^r_{23})(A^{r+1}_{21}+A^{r+1}_{32}).&(3.5c)\cr}$$
\advance\eqnumber by 1

By Eqs.~(2.6), one can further express $H$ in terms of the three sets
of boson operators.  For clarity, I use different notations
for these three sets of bosons,
$$ a\equiv a_1,\ a^+\equiv a^+_1;
    \ \ \ \ b\equiv a_3,\ b^+\equiv a_3^+;
    \ \ \ \ c\equiv a_2,\ c^+ \equiv a_2^+. \en$$
Now the Hamiltonian of Eq.~(3.4) can be written as
 $$  H = H_0 + V, \en $$
where $H_0$ contains only the quadratic terms and a constant,
 $$ \eqalign{
      H_0&=-{3\over8}N + \sum_{r=1}^{N/2}\biggl\{a^+_ra_r+b^+_rb_r+c^+_rc_r
      +{1\over4}(2J-1)\bigl[(a_r-b^+_r)(a^+_{r+1}-b_{r+1})\cr
        &+(a^+_r-b_r)(a_{r+1}-b^+_{r+1})
         +(c_r+c^+_r)(c_{r+1}+c^+_{r+1})\bigr]\biggr\},\cr} \en
 $$
and $V$ contains higher-order terms up to the sixth,
 $$ V = V_3 + V_4 + V_5 +V_6. \en$$

There is a close analogy between the bosonizations in the present
case and in Anderson's spin-wave theory [6], where the N\'eel state is
the vacuum state for the two sets of boson operators. Clearly,
$A_{00}$ corresponds to $s^z$ in the conventional spin-wave theory,
while $A_{n0}$ ($A_{0n}$) corresponds to $s^+$ ($s^-$), etc.
Notice that there are three sets of independent boson operators in the
present case, but there are only two in the conventional spin-wave
theory. This is because the symmetry-broken vacuum state (the N\'eel
state) in the conventional spin-wave theory is in the subspace of zero
$s_{\rm total}^z\ (\equiv \sum_l s^z_l)$ while the symmetry-broken
vacuum state (the dimer valence-bond state) in the present case is in
the subspace of zero vector $\vs_{\rm total}\ (\equiv \sum_l\vs_l)$.
It is already clear that one should expect a triplet
excitation of spin 1 for the present dimerized system.

After Fourier transformations for the three sets of boson operators,
 $$\eqalign{
	a_k&=\sqrt{2\over N}\sum_{r=1}^{N/2}\exp(-2ikr) a_r,\hskip.5truein
	b_k=\sqrt{2\over N}\sum_{r=1}^{N/2}\exp(-2ikr) b_r,\cr
	c_k&=\sqrt{2\over N}\sum_{r=1}^{N/1}\exp(-2ikr) c_r,\cr} \en
 $$
(the factor of 2 in the exponential functions is due to the
double spacings in the dimer index $r$,) $H_0$ of Eqs.~(3.8) can
be diagonalized by the usual Bogoliubov transformations,
 $$\eqalign{
	a_k&=\cosh\theta_k\alpha_k-\sinh\theta_k\beta^+_{-k},\hskip.5truein
	b_k=-\sinh\theta_k\alpha^+_{-k}+\cosh\theta_k\beta_k,\cr
	c_k&=\cosh\theta_k\gamma_k+\sinh\theta_k\gamma^+_{-k},\cr} \en
 $$
where $\theta_k$ is given by
 $$
	\tanh2\theta_k={(1-2J)\cos2k\over2-(1-2J)\cos2k}.\en
 $$

The diagonalized Hamiltonian $H_0$ can be simply written as
 $$
  H_0 = \sum_k\omega_k(\alpha^+_k\alpha_k+\beta^+_k\beta_k+\gamma^+_k
			\gamma_k)+E_0, \en
 $$
where the triplet spectrum is given by
 $$
	\omega_k=\sqrt{1-(1-2J)\cos2k}, \en
 $$
which agrees with Parkinson [7] at $J=0$, and where $E_0$
is defined as
 $$
	{E_0\over N}={3\over4}\sum_k\bigl[\sqrt{1-(1-2J)\cos2k}-1\bigr]
			-{3\over8}. \en
 $$
In Eqs.~(3.13) and (3.15) the summation over $k$ is defined as
 $$	\sum_k\equiv{1\over2\pi}\int^\pi_{-\pi}\,dk. \en $$

It is well known [18] that the ground state $\ket{\Phi_0}$ of
the quadratic Hamiltonian $H_0$ is given by the two-body form as
 $$ \ket{\Phi_0} = \exp\left[-\sum_k \tanh \theta_k \left(a^+_kb^+_{-k}
                -{1\over2}c^+_kc^+_{-k}\right)\right] \ket D, \en$$
where $\theta_k$ is determined by Eq.~(3.12) and $\ket D$ is the boson
vacuum state of Eq.~(3.2).

The ground-state energy $E_g$ within this spin-wave theory is given by
the expectation value of the full Hamiltonian of Eq.~(3.7) with
respect to $\ket{\Phi_0}$. Clearly, the odd-body terms yield zero, and
one has $ E_g = E_0 + \langle V_4
\rangle + \langle V_6\rangle$, where  $\langle V_4\rangle$ and
$\langle V_6\rangle$ can be calculated by Wick's theorem. If one
ignores $V_4$ and $V_6$ which represent spin-wave interactions, the
ground-state energy is then approximated by $E_0$ of Eq.~(3.15). This
is the result shown in Fig.~2, where I also include for comparison the
numerical results of Tonegawa and Harada [19], obtained by
extrapolating the finite-size calculations for $J<1/2$, and the exact
results by Parkinson [20] of the $N=20$ system for $J>1/2$. Notice
that at $J=1/2$ (the Majumdar-Ghosh point), Eq.~(3.15) gives the exact
result, namely $E_0/N=-3/8$.  This is not surprising because
$\ket{\Phi_0} = \ket D$ at this point. At $J=0$ (Heisenberg), one has
$E_0/N=-0.4498$, whereas the exact result by the Bethe ansatz [14] is
$-0.4432$. But one should be careful here because $J=0$ corresponds to
one of the terminating points at which the spin-wave theory is most
unreliable, as discussed below.

{}From the triplet spectrum of Eq.~(3.14) one sees that there are two
terminating points, $J=0$ and $1$, beyond which (i.e., $J<0$ and
$J>1$) the spin-wave excitations are unstable. In Fig.~3, the triplet
excitation spectrum is schematically shown for several values of the
coupling constant $J$. It is clear that in the region of $0<J<1$,
there is a nonzero gap, and this gap collapses at the terminating
points. In particular, the triplet spectrum is flat with the gap of 1
at $J=1/2$. The flatness reflects the fact that at $J=1/2$, $H_0$ of
Eq.~(3.8) contains no coupling at all between pairs of spins (dimers).
More realistic calculations for the excitations at $J=1/2$ were
performed by Shastry and Sutherland and others [21]. They obtained the
spectrum of soliton-like excitations with the minimum gap of 0.25 at
$k=0$ and $\pi$ and the maximum gap of 1 at $k=\pi/2$. Tonegawa and
Harada's numerical calculations [19] confirmed the nonzero gap at
$J=1/2$ and in the nearby region.  They predicted that the gap
collapses at $J\approx 0.3$, while Haldane [22], who used a fermion
representation, predicted this value to be about 1/6. This gapless
point seems to signal a phase transition from the dimerized phase to a
critical phase similar to the Heisenberg model at $J=0$.  In any case,
the triplet spectrum of $\sqrt2\sin k$ from Eq.~(3.14) at $J=0$ seems
to agree well with the exact result of ${\pi\over2}\sin k$, as pointed
out by Parkinson [7].

A more intriguing situation occurs for $J>1/2$, where the spin-wave
spectrum has a minimum at $k=\pi/2$. In particular, at $J=1$, the
spectrum is gapless with a cusp at $k=\pi/2$. Whether or not this
suggests a phase change in the spatial periodicity of the system from
double to four-fold, for example, is still unclear. The numerical
calculations of the structure factor by Tonegawa and Harada [19]
certainly showed a complicated feature for $J>1/2$. There is also
numerical evidence in the excitation spectra, which suggests that the
spatial periodicity is not two-fold in the region near $J=1$ [20].
Clearly, the spin-wave theory described here is not adequate for this
task and higher-order calculations are needed.

One can also straightforwardly calculate the long-range dimerization
order parameter within the present spin-wave theory. The dimer order
parameter $D$ is defined as
$$ D\equiv \langle\vs_{l-1}\cdot\vs_l\rangle
-\langle\vs_l\cdot\vs_{l+1}\rangle
=\langle\vs_1(r)\cdot\vs_2(r)\rangle
-\langle\vs_2(r)\cdot\vs_1(r+1)\rangle, \en$$
where the expectation is with respect to the ground state of the
system. By Eqs.~(2.5) and (2.6), using the spin-wave ground state
$\ket{\Phi_0}$ of Eq.~(3.17), it is easy to show that in the spin-wave
approximation, $D$ is nonzero in the region $0<J<1$ and gradually
diminishes when $J$ moves toward the two terminating points. But at
the terminating points ($J=0,1$), $D$ diverges to $-\infty$, implying a
breakdown of the spin-wave theory. In the following section, I
provide an alternative approach to the dimerization problem by
applying the microscopic CCM.

\vskip \baselineskip

\newcount\eqnumber
\eqnumber=0
\def\en{\global\advance\eqnumber by 1
          \eqno(4.\the\eqnumber)}

\centerline{\bf IV. The Coupled-Cluster Approach}
\nobreak\vskip \baselineskip \nobreak

The CCM is widely recognized nowadays as providing one of the most
universally applicable, most powerful and most accurate of all
microscopic formulations of quantum many-body theory [13]. The recent
application of the CCM to various spin models has produced excellent
numerical results [14,15]. It therefore seems very appropriate and
timely to apply the CCM to the dimerization problem. The interested
reader is referred to Ref.~[13] for the general formalism of the CCM
and to Ref.~[14] for its particular application to the spin systems
with an anticipated Ising-like long-range order.

Generally speaking, the CCM starts with a proper model state
$\ket\Phi$, which is usually a simple, uncorrelated many-body
wavefunction, and incorporates many-body correlations on top of
$\ket\Phi$ by acting on it with an exponentiated correlation operator
$S$. This operator $S$ consists of purely so-called configuration
creation operators with respect to the model state $\ket\Phi$, and
is partitioned by one-body, two-body, ..., up to $N$-body
correlation operators with $N$ the number of particles in the system.
Thus, the CCM {\it ansatz} for the ground ket state is
$$ \ket{\Psi_g} = {\rm e}^S\ket\Phi. \en$$
The Schr\"odinger equation of the ground state, after a simple
manipulation, can then be written as
$$ {\rm e}^{-S}H{\rm e}^S\ket\Phi=E_g\ket\Phi, \en$$
where $E_g$ is the ground-state energy, and where the
similarity-transformed Hamiltonian can be expressed as a series of
nested commutators,
$$ {\rm e}^{-S}H{\rm e}^S=H+[H,S]+{1\over2!}[[H,S],S]+\cdots, \en$$
which usually terminates at the fourth-order for most Hamiltonians
with pair-interaction potentials [13]. Eqs.~(4.1)-(4.3) are the
hallmarks of the CCM.

For the case of dimerization under consideration, it is natural to
choose the dimer state $\ket D$ of Eq.~(3.2) as the model state,
namely $\ket\Phi = \ket D$. The configuration creation operators with
respect to this model state $\ket D$ are clearly given by any
combinations of the three operators $A^r_{10}$, $A^r_{20}$, and
$A^r_{30}$. Their hermitian conjugates are the corresponding
destruction operators. Since the antiferromagnetic ground state is
definitely in the sector of zero $s^z_{\rm total}$, the correlation
operator $S$ is in general written in the form
$$ S=\sum_{n=1}^{N/2} S_n, \en$$
where $N/2$ is the total number of valence bonds of the spin-$1\over2$
chain, and the $n$-body correlation operators $S_n$ are given respectively
by
$$ \eqalignno{
  S_1&\equiv \sum_{r=1}^{N/2}\cs_r A^r_{20}, &(4.5a)\cr
  S_2&\equiv \mathop{{\sum}'}_{r,r'}^{N/2}\left[
   \cs^{(1)}_{r,r'}A^r_{10}A^{r'}_{30}
   -{1\over2!}\cs^{(2)}_{r,r'}A^r_{20}A^{r'}_{20}\right],&(4.5b)\cr
  S_3&\equiv \mathop{{\sum}'}_{r,r',r''}^{N/2}\left[
   \cs^{(1)}_{r,r',r''}A^r_{10}A^{r'}_{30}A^{r''}_{20}
   -{1\over3!}\cs^{(2)}_{r,r',r''}A^r_{20}A^{r'}_{20}A^{r''}_{20}\right],
   &(4.5c) \cr}$$
\advance\eqnumber by 1
etc. In Eq.~(4.5b) and (4.5c) the primes on the summations imply
exclusion of the terms with any pair of indices being equal. We notice
the similarity between the spin-wave ground state $\ket{\Phi_0}$ of
Eq.~(3.17) and the CCM state $\ket{\Psi_g}$ of Eq.~(4.1) if $S$ is
replaced by $S_2$. W also notice the similarity between the present
CCM analysis and that of Ref.~15 where the spin-1 model state is given
by the simple planar configuration with $s^z_l=0$ for all sites $l$.

The ground-state energy is obtained by taking the inner product of the
Schr\"odinger equation (4.2) with the model state $\ket D$ itself,
namely
$$ E_g=\bra D {\rm e}^{-S}H{\rm e}^S\ket D; \en$$
the correlation coefficients $\{\cs_{r,r',...}\}$ are determined by
the coupled set of equations obtained by taking inner products of
Eq.~(4.2) with states constructed from the corresponding
destruction operators, namely
$$ \bra D A^r_{02}{\rm e}^{-S}H{\rm e}^S\ket D = 0, \ \ \ \
   {\rm for\ all\ }r , \en$$
for the one-body equation;
$$ \bra D A^r_{01}A^{r'}_{03}{\rm e}^{-S}H{\rm e}^S\ket D=0,
   \ \ \ \ {\rm for\ all\ }r,r'(\not=r) \en$$
and
$$ \bra D A^r_{02}A^{r'}_{02}{\rm e}^{-S}H{\rm e}^S\ket D = 0,
   \ \ \ \ {\rm for\ all\ }r,r'(\not=r) \en$$
for the two-body equations. The three-body equations and higher-order
many-body equations are obtained in a similar fashion.

One sees that the similarity-transformed Hamiltonian of Eq.~(4.3) is
needed in all of the above equations. I leave details of the derivation
to the Appendix, and only point out that, as expected, the otherwise
infinite expansion series of Eq.~(4.3) indeed terminates at the
fourth-order term. In fact, the exact energy equation (4.6) can be
straightforwardly derived as
$$ {E_g\over N}={1\over8}[(1-2J)(2b_1^{(1)}+b_1^{(2)}-a)-3], \en$$
where I have used the translational and reflectional symmetries,
setting accordingly,
$$ \cs_r=a, \ \ \ \ \cs^{(i)}_{r_1,r_2}=\cs^{(i)}_{r_2,r_1}=b_r^{(i)},
   \ \ {\rm with\ } i=1,2 \ {\rm and\ } r=r_2-r_1, \en$$
It is also obvious that $b_{-r}^{(i)}=b_r^{(i)}$.

The exact one-body equation of Eq.~(4.7) can also be easily derived.
It couples only to the two-body coefficients. Similarly, the two-body
equations of Eqs.~(4.8) and (4.9) couple only to the one-body and the
three-body coefficients, and so on.  From the one-body equation, it is
interesting to note that the physical solution is given by $a=0$,
implying no one-body correlations for the dimerization
problem. This is not surprising because the model state $\ket D$ is in
the sector of zero total spin vector (i.e., $\vs_{\rm total}=0$), and
the one-body correlation operator $S_1$ of Eq.~(4.5a) will take the state
out of this sector. Furthermore, if one assumes that
the two sets of the two-body
correlation coefficients are identical, namely
$$ b^{(1)}_r = b^{(2)}_r, \en$$
the two-body correlation operator $S_2$ then commmutes with the total
spin vector $\vs_{\rm total}$. This is a necessary condition if one
requires the CCM ground state to be in the sector of zero
$\vs_{\rm total}$. (Actually, one requires every correlation operator
$S_n$ to commute with $\vs_{\rm total}$.)  The energy equation is now
reduced to
$$ {E_g\over N}={3\over8}[(1-2J)b_1-1]. \en$$

One clearly needs to employ an approximation scheme for any practical
calculation. The most common approximation scheme in the CCM is the
so-called SUB$n$ scheme, in which one keeps up to $n$-body correlation
operators and sets all the higher-order many-body correlation
operators $S_m\ (m>n)$ to zero. I consider the SUB2 scheme here. I
find that within the SUB2 scheme, the condition of Eq.~(4.12) is
indeed satisfied. After simplification, the two identical
two-body equations are given by
$$ {1\over2}\sum_{\rho=\pm1}(K_3\delta_{r\rho}+K_2b_r-2K_1b_{r+\rho}
   +K_1\sum_{r'\not=0}b_{r'}b_{r+\rho-r'}) = 0,\ \ \ \ r\not=0 \en$$
and
$$  K_1\equiv 1-2J, \ \ \ \
    K_2\equiv 4(1-2K_1b_1), \ \ \ \
    K_3\equiv K_1(1+4b_1^2)-2(1+2J)b_1. \en $$

A simpler approximation can be made from the full SUB2 equation
(4.14), namely the so-called SUB2-2 scheme in which one keeps only the
single coefficient, $b_1$, setting all other $b_r = 0\ (\vert r\vert >1)$.
Eq.~(4.14) then reduces to
$$ 1-2J+2(3-2J)b_1-9(1-2J)b_1^2=0, \en$$
with the physical solution
$$ b_1={1\over9(1-2J)}[3-2J-\sqrt{40J^2-48J+18}]. \en$$

The full SUB2 equation (4.14) can also be solved by a Fourier
transformation exactly similar to Eq.~(3.9) of Ref.~[15]. Here I only
quote the final result given by the following self-consistency
equation for $b_1$,
$$ b_1={1\over3K_1}\left(2-{K_2\over2}{1\over2\pi}\int_{-\pi}^\pi
    dq\,\sqrt{1-k_1\cos2q+k_2\cos^22q}\right), \en$$
where the constants $k_1$ and $k_2$ are defined by
$$ k_1\equiv {1\over K_2^2}(4K_1K_2+8K_1^2b_1-4K_1^2X), \ \ \ \
   k_2\equiv {4K_1(K_1-K_3)\over K_2^2},      \eqno(4.19a)$$
and where $X$ is defined by
$$ \eqalignno{
   X&\equiv\sum_{r=1}^{N/2}b_rb_{r+1}\cr
    &={1\over2\pi}\int^\pi_{-\pi}dq\,{1\over4K_1^2\cos2q}\cr
    &\times\left(2K_1\cos2q-K_2+K_2\sqrt{1-k_1\cos2q+k_2\cos^22q}\right)^2.
           &(4.19b)\cr}$$
\advance\eqnumber by 1

After obtaining $b_1$ as a function of $J$ from Eq.~(4.17) or (4.18),
the ground-state energy is then given by substituting $b_1$ into
Eq.~(4.13). These ground-state energies are shown in Fig.~2, together
with the results of the spin-wave theory and of the numerical
calculations [19,20] for comparison. Similar to the spin-wave theory
described in Sec.~III, at $J=1/2$ (the Majumdar-Ghosh point), the
exact result is recovered for both the SUB2-2 and full SUB2 schemes,
namely, $b_1 = 0$ and $E_g/N=-3/8$. At the Heisenberg point ($J=0$),
the SUB2-2 and full SUB2 schemes give $E_g/N = -0.4268, -0.4298$
respectively, slightly higher than the exact result of $-0.4432$.
Furthermore, it is interesting to observe that in the full SUB2
scheme, there are also two terminating points, $J_c^{(1)}=-0.4443$ and
$J_c^{(2)}=1.591$, beyond which, namely for $J < J_c^{(1)}$ and $J >
J_c^{(2)}$, there is no real solution in Eq.~(4.18). The corresponding
energy values are $-0.5172$ and $-0.6977$ respectively.  The CCM SUB2
terminating points have been argued to correspond to the phase
transition critical points in the past [14,15]. It seems reasonable to
consider this possibility again here. From Fig.~2, one sees that the
extremely simple SUB2 scheme gives much better results for a wide
range of the coupling constant $J$ than the spin-wave theory does, at
least as far as the ground-state energy is concerned.

\vskip \baselineskip

\centerline{\bf V. Discussion}
\nobreak\vskip\baselineskip\nobreak

In this paper, I have studied the dimerization problem by a
microscopic approach, employing the proper set of composite operators
of Parkinson [7]. Two approximations schemes, namely the spin-wave
theory and a CCM analysis, have been applied to the 1D frustrated
spin-$1\over2$ model. The ground-state and low-lying excited energies
are obtained as functions of the coupling constant. The implications
of possible phase transitions at the naturally arising terminating
points of the solutions have been discussed. Another approach may be
provided by a variational trial wave function of the type of
Eq.~(4.1), similar to the calculation of Sachdev for the
spin-$1\over2$ Heisenberg model [23].

{}From the present preliminary attempt to formulate a microscopic theory
for the dimerization problem, it is clear that higher-order
calculations within the present analysis are needed for both the
ground and excited states. The very successful applications [14] of
the CCM to the spin systems with an Ising-like long-range order seem
to suggest that the CCM can also provide a systematic and potentially
accurate approximation scheme for the dimerization problem.
Furthermore, within the formalism presented in this paper, it is
straightforward to extend the same analysis to both higher-order
dimensionality and/or spin systems with spin quantum number greater
than one half. In particular, the 2D spin-$1\over2$ Heisenberg model
on the square lattice with $J_1$-$J_2$ couplings [9,11] has been under
intensive study for its possible dimerization. An equally interesting
Hamiltonian model is provided by the 1D spin-1 Heisenberg-biquadratic
systems, where it is known the ground state is dimerized at a
particular coupling constant [4], and where trimerization is also
possible in another region [10].

\vskip \baselineskip

\centerline{\bf Acknowledgements}
\nobreak\vskip\baselineskip\nobreak

I am grateful to J.B. Parkinson for drawing my attention to Ref.~[7]
which forms the basis of this work and for providing some numerical
results of finite size calculations prior to publication, and to R.F.
Bishop for suggesting application of the CCM to this problem. I also
thank C.E. Campbell, N.J.  Davidson, C. Zeng, and D.J.J. Farnell for
many useful discussions.

\vskip \baselineskip

\newcount\eqnumber
\eqnumber=0
\def\en{\global\advance\eqnumber by 1
          \eqno(A.\the\eqnumber)}

\centerline{\bf Appendix}
\nobreak\vskip \baselineskip \nobreak

In this appendix, I derive the similarity transformations within the
SUB1 and SUB2 schemes of the CCM, described in Sec.~IV.

Notice that in the similarity transformation of Eq.~(4.3), any
quadratic term in the Hamiltonian of Eqs.~(3.4)-(3.5) can be
transformed as
$$ {\rm e}^{-S}A^r_{ij}A^{r+1}_{kl}{\rm e}^{S} = ({\rm
e}^{-S}A^r_{ij}{\rm e}^{S}) ({\rm e}^{-S}A^{r+1}_{kl}{\rm e}^{S}),
  \en$$
and each similarity-transformed operator can be expanded as a
series of nested-commutators,
$$ {\rm e}^{-S}A^r_{ij}{\rm e}^{S} =
A^r_{ij} +[A^r_{ij}, S] +{1\over2!}[[A^r_{ij}, S],S] + \cdots. \en$$
Since the correlation operator $S$ consists of only the creation
operators $A_{10}$, $A_{30}$ and $A_{20}$, the expansion series of
Eq.~(A.2) terminates at most at the second-order by the pseudo-spin
algebra of Eq.~(2.4).  Therefore the similarity-transformed
Hamiltonian of Eq.~(4.3) terminates at the fourth-order.

In the SUB1 scheme, one replaces $S\rightarrow S_1$, where $S_1$ is
given by Eq.~(4.5a). From Eq.~(A.2), it is straightforward to derive
the following SUB1 similarity transformations
$$  \eqalign{
 \tilde A^r_{n0} &= A^r_{n0}, \ \ \ \  \tilde A^r_{nm} = A^r_{nm},\cr
 \tilde A^r_{00} &= A^r_{00}-\cs_r A^r_{20},\ \ \ \
     \tilde A^r_{n2} = A^r_{n2} + \cs_r A^r_{n0}, \cr
 \tilde A^r_{02} &= A^r_{02} + \cs_r (A^r_{00}-A^r_{22}) -
      \cs^2_r A^r_{20}, \cr} \en $$
where $n=1,2,3$ and $m=1,3$, and where the definition
$$ \tilde A^r_{ij} \equiv {\rm e}^{-S_1}A^r_{ij}{\rm e}^{S_1}, \en$$
is used.

In the SUB2 scheme, one replaces $S\rightarrow S_1+S_2$. One can
firstly make the SUB1 similarity transform by Eq.~(A.3), and then apply
the following SUB2 similarity transform for each $A^r_{ij}$,
$$ \eqalignno{
  \bar A^r_{n0} &= A^r_{n0}, &(A.5a)\cr
  \bar A^r_{n1} &= A^r_{n1} + \mathop{{\sum}'}_{r'} \cs^{(1)}_{r,r'}
         A^r_{n0}A^{r'}_{30},&(A.5b)\cr
  \bar A^r_{n2} &= A^r_{n2} - \mathop{{\sum}'}_{r'} \cs^{(2)}_{r,r'}
        A^r_{n0}A^{r'}_{20}, &(A.5c)\cr
  \bar A^r_{n1} &= A^r_{n3} + \mathop{{\sum}'}_{r'} \cs^{(1)}_{r',r}
        A^r_{n0}A^{r'}_{10}, &(A.5d)\cr
  \bar A^r_{00} &= A^r_{00} -\mathop{{\sum}'}_{r'}
       (\cs^{(1)}_{r,r'}A^r_{10}A^{r'}_{30}
       +\cs^{(1)}_{r',r}A^{r'}_{10}A^r_{30}
       -\cs^{(2)}_{r,r'}A^r_{20}A^{r'}_{20}), &(A.5e)\cr
  \bar A^r_{01}
 &= A^r_{01} +
   \mathop{{\sum}'}_{r'}\bigl(
         \cs^{(1)}_{r,r'}A^{r'}_{30}(A^r_{00}-A^r_{11})
        -\cs^{(1)}_{r',r}A^{r'}_{10}A^r_{31}
        +\cs^{(2)}_{r,r'}A^{r'}_{20}A^r_{21}
           \bigr)\cr
 &+{1\over2}\mathop{{\sum}'}_{r',r''}\cs^{(1)}_{r,r'}A^{r'}_{30}
   \bigl(
       \cs^{(2)}_{r,r''}A^r_{20}A^{r''}_{20}
     -2\cs^{(1)}_{r,r''}A^r_{10}A^{r''}_{30}
     - \cs^{(1)}_{r'',r}A^{r''}_{10}A^r_{30}
   \bigr)\cr
 &-{1\over2}\mathop{{\sum}'}_{r',r''}\bigl(
    \cs^{(1)}_{r',r}\cs^{(1)}_{r,r''}A^{r'}_{10}A^r_{30}A^{r''}_{30}
   -\cs^{(2)}_{r,r'}\cs^{(1)}_{r,r''}A^{r'}_{20}A^r_{20}A^{r''}_{30}
                         \bigr),&(A.5f)\cr
 \bar A^r_{02}
  &= A^r_{02} - \mathop{{\sum}'}_{r'}
    \bigl(
       \cs^{(2)}_{r,r'}A^{r'}_{20}(A^r_{00}-A^r_{22})
      +\cs^{(1)}_{r',r}A^{r'}_{10}A^r_{32}
      +\cs^{(1)}_{r,r'}A^{r'}_{30}A^r_{12}
     \bigr)\cr
   &-{1\over2}\mathop{{\sum}'}_{r',r''}\cs^{(2)}_{r,r'}A^{r'}_{20}
    \bigl(
     2\cs^{(2)}_{r,r''}A^r_{20}A^{r''}_{20}
     -\cs^{(1)}_{r,r''}A^r_{10}A^{r''}_{30}
     -\cs^{(1)}_{r'',r}A^{r''}_{10}A^r_{30}
    \bigr)\cr
   &+{1\over2}\mathop{{\sum}'}_{r',r''}\bigl(
     \cs^{(1)}_{r',r}\cs^{(2)}_{r,r''}A^{r'}_{10}A^r_{30}A^{r''}_{20}
    +\cs^{(1)}_{r,r''}\cs^{(2)}_{r,r'}A^{r''}_{30}A^r_{10}A^{r'}_{20}
    \bigr), &(A.5g)\cr
 \bar A^r_{03}
 &= A^r_{03} +
   \mathop{{\sum}'}_{r'}\bigl(
         \cs^{(1)}_{r',r}A^{r'}_{10}(A^r_{00}-A^r_{33})
        -\cs^{(1)}_{r,r'}A^{r'}_{30}A^r_{13}
        +\cs^{(2)}_{r,r'}A^{r'}_{20}A^r_{23}
           \bigr)\cr
 &+{1\over2}\mathop{{\sum}'}_{r',r''}\cs^{(1)}_{r',r}A^{r'}_{10}
   \bigl(
       \cs^{(2)}_{r,r''}A^r_{20}A^{r''}_{20}
     -2\cs^{(1)}_{r'',r}A^r_{30}A^{r''}_{10}
     - \cs^{(1)}_{r,r''}A^{r''}_{30}A^r_{10}
   \bigr)\cr
 &-{1\over2}\mathop{{\sum}'}_{r',r''}\bigl(
    \cs^{(1)}_{r,r'}\cs^{(1)}_{r'',r}A^{r'}_{30}A^r_{10}A^{r''}_{10}
   -\cs^{(2)}_{r,r'}\cs^{(1)}_{r'',r}A^{r'}_{20}A^r_{20}A^{r''}_{10}
                         \bigr), &(A.5h)\cr}$$
\advance\eqnumber by 1
where $n=1,2,3$, and the primes on the summations imply exclusion of
any pair of indices being equal, and where the operators with a bar
represents the similarity transform for the $S_2$ correlation operator,
$$ \bar A^r_{ij} \equiv {\rm e}^{-S_2}A^r_{ij}{\rm e}^{S_2}. \en$$
In deriving  Eqs.~(A.5), I have used the fact that
$$ \cs^{(1)}_{r,r}=\cs^{(2)}_{r,r}=0, \ \ \ \
   \cs^{(2)}_{r,r'}=\cs^{(2)}_{r',r}. \en$$
After the similarity transformations of Eqs.~(A.3) and (A.5), the CCM
equations of Eqs.~(4.6)-(4.9) can be derived by using the pseudo-spin
algebra of Eq.~(2.4) to move in each term all the creation
operators $A_{10},A_{20},A_{30}$ to the left, all the detruction
operators $A_{01},A_{02},A_{03}$ (also $A_{nm},\ n,m\not=0$) to the
right, and by using the fact that
$$ \bra D A_{n0}^r = A_{mn}^r\ket D = 0, \ \ \ \ n\not=0, \en$$
and
$$ \bra D A_{00}^r = \bra D, \ \ \ \ A_{00}^r\ket D = \ket D, \en$$
for all $r$.

\vskip \baselineskip

\newcount\eqnumber
\eqnumber=0
\def\refno{\global\advance\eqnumber by 1  \the\eqnumber}

\centerline{\bf References} \nobreak \vskip.3truein \nobreak

\item{[\refno]}C.K. Majumdar and D.K. Ghosh, J. Phys. C {\bf 3}, 91
(1970); J. Math. Phys. {\bf 10}, 1388, 1399 (1969).

\item{[\refno]}I. Affleck, T. Kennedy, E. Lieb, and H. Tasaki, Phys.
Rev. Lett. {\bf 59}, 799 (1987).

\item{[\refno]}J.B. Parkinson, J. Phys. C {\bf 20}, 21 (1988); M.N.
Barber and M.T. Batchelor, Phys. Rev. B {\bf 40}, 4621 (1989); A.
Kl\"umper, Europhys. Lett. {\bf 9}, 815 (1989); I. Affleck, J. Phys.:
Condens. Matter {\bf 2}, 405 (1990).

\item{[\refno]}Y. Xian, Phys. Lett. A {\bf 183}, 437 (1993).

\item{[\refno]}``Valence-bond lattices'' is used in this article
in order to distinguish from the well-known ``valence-bond solids''
defined in Ref.~[2] as a single valence-bond configuration of a
many-spin system, including the nondegenerate spin-liquid state which
has no ordinary symmetry breaking.

\item{[\refno]}P.W. Anderson, Phys. Rev. {\bf 86}, 694 (1952); R.
Kubo, {\it ibid.} {\bf 87}, 568 (1952); T. Oguchi, {\it ibid.} {\bf
117}, 117 (1960).

\item{[\refno]}J.B. Parkinson, J. Phys. C {\bf12}, 2873 (1979).

\item{[\refno]}A.V. Chubukov, Phys. Rev. B {\bf 43}, 3337 (1991).

\item{[\refno]}N. Read and S. Sachdev, Phys. Rev. Lett. {\bf 62}, 1694
(1989); Phys. Rev. B {\bf 42}, 4568 (1990).

\item{[\refno]}Y. Xian, J. Phys.: Condens. Matter {\bf 5}, 7489 (1993).

\item{[\refno]}R.R.P. Singh and R. Narayanan, Phys. Rev. Lett. {\bf
65}, 1072 (1990); A.V. Chubukov and Th. Jolicoeur, Phys. Rev. B {\bf 44},
12050 (1991).

\item{[\refno]}F.J. Dyson, Phys. Rev. {\bf 102}, 1217, 1230 (1956);
S.V. Mal\'eev Sov. Phys. JETP {\bf 6}, 776 (1958). For the
pseudospin operators in the present context, their Holstein-Primakoff
transformations were given by L.R. Mead and N. Papanicolaou in Phys.
Rev. B {\bf 28}, 1633 (1983).

\item{[\refno]}R.F. Bishop and H. K\"ummel, Phys. Today {\bf 40(3)},
52 (1987); R.F. Bishop, Theo. Chim. Acta {\bf 80}, 95 (1991).

\item{[\refno]}R.F. Bishop, J.B. Parkinson, and Y. Xian, Phys. Rev. B
{\bf 43}, 13782 (1991); {\bf 44}, 9425 (1991); J. Phys.: Condens.
Matter {\bf 4}, 5783 (1992).

\item{[\refno]}R.F. Bishop, J.B. Parkinson, and Y. Xian, J. Phys.:
Condens. Matter {\bf 5}, 9169 (1993).

\item{[\refno]}M. den Nijs and K. Rommelse, Phys. Rev. B {\bf 40},
4709 (1989); G. Santoro and M. Fabrizio, ``Disordered flat phase in
a solid on solid model of fcc(110) surface and dimer states in quantum
spin-$1\over2$ chains,'' preprint (1994).

\item{[\refno]}In Ref.~[7] the spin operators $S^+$ and $S^-$ are
interchanged when compared with Eqs.~(2.5) in this paper.

\item{[\refno]}J.P. Blaizot and G. Ripka, {\it Quantum Theory of
Finite Systems}, MIT, London, 1986.

\item{[\refno]}T. Tonegawa and I. Harada, J. Phys. Soc. Japan {\bf
56}, 2153 (1987).

\item{[\refno]}Courtesy of J.B. Parkinson.

\item{[\refno]}S.S. Shastry and B. Sutherland, Phys. Rev. Lett. {\bf
47}, 964 (1981); W.J. Caspers, K.M. Emmett, and W. Magnus, J. Phys. A
{\bf 17}, 2697 (1984).

\item{[\refno]}F.D.M. Haldane, Phys. Rev. B {\bf 25}, 4925 (1982).

\item{[\refno]}S. Sachdev, Phys. Rev. B {\bf 39}, 12232 (1989).

\vfill
\eject

\centerline{\bf Figure Captions}
\vskip \baselineskip

\item{Fig. 1} The perfect dimer state and dimer indexing. Each bond
represents a singlet configuration as given by Eq.~(2.1).

\vskip\baselineskip

\item{Fig.~2} Ground-state energy per spin as a function of the
coupling constant $J$. Shown are results from the spin-wave theory
(dotted), the SUB2-2 scheme (short dashed), and the full SUB2 scheme
(long dashed). The terminating points of the SUB2 scheme and spin-wave
theory are indicated.  The numerical results from Ref.~[19,20]
are also included (solid).

\vskip\baselineskip

\item{Fig.~3} Schematic plots of the triplet excitation spectrum of
Eq.~(3.14) for various values of the coupling constant $J$.

\bye